\documentclass[english,aip,pof,twocolumn,tightenlines]{revtex4-1}
\usepackage[T1]{fontenc}
\usepackage[latin9]{inputenc}
\usepackage{geometry}
\usepackage[active]{srcltx}
\usepackage{amsmath}

\makeatletter
\def\vec#1{\boldsymbol{#1}}

\usepackage[mathlines]{lineno}

\makeatother

\usepackage{babel}
\begin{document}
\title{Comment on \textquotedblleft Linear stability of a rotating channel
flow subjected to a static magnetic field \textquotedblright{} {[}Phys.
Fluids 34, 054116 (2022){]}}
\author{J\={a}nis Priede}
\altaffiliation[Also at ]{Fluid and Complex Systems Research Centre, Coventry University, 
Coventry, CV1 5FB, United Kingdom }

\affiliation{Department of Physics, University of Latvia, Riga, Latvia}
\email{j.priede@coventry.ac.uk}

\selectlanguage{english}%
\begin{abstract}
Recently, Sengupta and Ghosh {[}Phys. Fluids \textbf{34}, 054116,
(2022){]} published a linear stability analysis of a pressure-driven
channel flow which is subject to rotation around a spanwise axis and
a uniform magnetic field applied in the same direction. Unfortunately,
the formulation of the magnetohydrodynamic part of the problem contains
an elementary error which makes the obtained results unphysical. The
error is due to unfounded omission of the electric potential contribution
in the induced electric current which, thus, does not satisfy the
charge conservation.
\end{abstract}
\maketitle
Fluid moving with velocity $\vec{v}$ in the magnetic field $\vec{B}$
experiences the electromotive force (emf) $\vec{v}\times\vec{B}.$
If the fluid is electrically conducting, this emf can drive an electric
current. In general, the induced emf can have a non-zero divergence.
If this is the case, not all of the free charge carriers can follow
such an emf; some get stuck in the fluid so creating a non-zero volumetric
charge distribution. \citep{Davidson2016} The latter gives rise to
the static electric field $\vec{E}=-\vec{\nabla}\phi,$ which complements
the flow-induced emf in Ohm's law for moving medium so that the induced
current density,

\[
\vec{j}=\sigma(-\vec{\nabla}\phi+\vec{v}\times\vec{B}),
\]
can satisfy the charge conservation constraint $\vec{\nabla}\cdot\vec{j}=0;$
here $\sigma$ stands for the electrical conductivity of the fluid.
Although these equations appear in the paper \citep{Sengupta2022}
(the latter as Eq. (3) and the former a few lines further, where the
authors incorrectly refer to $\sigma$ as the ``charge density''),
the potential is omitted afterwards without any justification. This
omission becomes apparent in Eqs. (5,6) where the non-dimensionalized
$x$ and $y$ components of the Lorentz force $\vec{f}=\vec{j}\times\vec{B},$
which are multiplied with $\mathit{Ha}^{2}$ -- the Hartmann number
squared, can be seen to contain only the respective velocity components
but no electric potential. In the following, we show that the omitted
terms are in general comparable to the retained ones. Therefore, this
omission is not an approximation but rather an error which makes the
obtained results unphysical.

The unphysicality of this omission crops up first in the base flow,
which is invariant along the magnetic field: $(\vec{B}\cdot\vec{\nabla})\vec{v}=0.$
Since the emf induced by such a flow is irrotational: $\vec{\nabla}\times(\vec{v}\times\vec{B})=0,$
so is also the induced current. It means that the induced current
cannot close entirely in the fluid; electrically conducting (and interconnected)
walls are necessary for this. Unfortunately, the authors do not specify
electrical properties of the walls or any boundary conditions for
the electric potential.

If the current cannot close, the charge piles up in the fluid producing
a potential gradient which eventually cancels out the irrotational
emf. If there is no current, there is no Lorentz force either. Thus,
the flow remains unchanged. The authors, however, find that the increase
of the magnetic field, which is spanwise, results in the non-magnetic
Poiseuille flow morphing into its magnetohydrodynamic counterpart
-- the Hartmann flow. But it is important to note that the genuine
Hartmann flow requires a transverse magnetic field. \citep{Davidson2016}

Alternatively, if the walls are electrically conducting and interconnected
so that a horizontally uniform background current can flow across
the channel, the charge conservation would require this current to
have a spatially constant density. This, in turn, would result in
a constant electromagnetic braking force which would just modify the
applied pressure gradient but not the velocity distribution. Evidently,
the alteration of the base flow profile found by the authors is an
unphysical effect due to the violation of the charge conservation
caused by the spurious non-uniformity of the induced current which
flows across the channel. 

The same conclusion follows also from more general considerations
which are analogous to those discussed above with regard to the emf.
Namely, if the electric current distribution is invariant along the
magnetic field: $(\vec{B}\cdot\vec{\nabla})\vec{j}=0,$ the resulting
Lorentz force is irrotational: $\vec{\nabla}\times(\vec{j}\times\vec{B})=0.$
Hence, it can affect only the pressure distribution but not the circulation
of fluid.  It implies that the effect of the magnetic field has to
vanish when the flow disturbances are aligned with the field corresponding
to the spanwise wave number $\beta=0.$ Unfortunately, this again
is not the case in Eq. (12) of \citet{Sengupta2022} where the term
with $\mathit{Ha}^{2}$ can be seen to persist in this limit. 

In fact, if the flow is invariant along the magnetic field and has
closed streamlines, i.e., a zero total flow rate, no current is induced
regardless of electrical conductivity of the walls. In this case,
the induced electric potential, which can be written explicitly in
terms of the Lagrange stream function $\psi$ with $\vec{v}=\vec{\nabla}\times(\vec{e}_{B}\psi)$
and $\vec{e}_{B}=\vec{B}/B$ as $\phi=-B\psi,$ not only compensates
the emf but also satisfies the boundary conditions for perfectly conducting
$(\left.\phi\right|_{s}=0)$ as well as insulating $(\left.\partial_{n}\phi\right|_{s}=0)$
walls because analogous hydrodynamic boundary conditions are satisfied
by $\psi.$

\section*{Author declarations}

\subsection*{Conflict of Interest}

The author has no conflicts to disclose.

\bibliographystyle{aipnum4-1}
\bibliography{aip_phf34}

\end{document}